# Is Anyone Out There? Unpacking Q&A Hashtags on Twitter

**Jeffrey M. Rzeszotarski[1,2], Emma S. Spiro[1,3], Jorge Nathan Matias[1,4],
Andrés Monroy-Hernández[1], Meredith Ringel Morris[1]**

[1]Microsoft Research, Redmond, WA, USA
[2]Human-Computer Interaction Institute, Carnegie Mellon University, Pittsburgh, PA, USA
[3]Information School, University of Washington, Seattle, WA, USA
[4]MIT Media Lab, Massachusetts Institute of Technology, Cambridge, MA, USA
jeffrz@cs.cmu.edu, espiro@uw.edu, jnmatias@mit.edu, {andresmh, merrie}@microsoft.com

**ABSTRACT**
In addition to posting news and status updates, many Twitter users post questions that seek various types of subjective and objective information. These questions are often labeled with "Q&A" hashtags, such as *#lazyweb* or *#twoogle*. We surveyed Twitter users and found they employ these Q&A hashtags both as a topical signifier (this tweet needs an answer!) and to reach out to those beyond their immediate followers (a community of helpful tweeters who monitor the hashtag). However, our log analysis of thousands of hashtagged Q&A exchanges reveals that nearly all replies to hashtagged questions come from a user's immediate follower network, contradicting users' beliefs that they are tapping into a larger community by tagging their question tweets. This finding has implications for designing next-generation social search systems that reach and engage a wide audience of answerers.

**Author Keywords**
Twitter; social search; Q&A; information seeking; hashtags

**ACM Classification Keywords**
H.5.m. Information interfaces and presentation: Misc.

## INTRODUCTION AND RELATED WORK
Social networking sites (SNSes) have the potential to provide immense reach to their users. People are able to connect with not only their close friends, but also friends separated by large distances, unfamiliar acquaintances, and even complete strangers. This unprecedented ability to interact with people far and wide has changed not only the way we interact with others, but also how we consume information. Now, people are free to leverage online connections as part of the information-seeking process. By harnessing social media, a person can easily access friends who may have dealt with similar questions in the past, or reach strangers who might be domain experts on a topic. A substantial portion of SNS users have engaged in harnessing these new media to ask a variety of factual and subjective questions, including on public SNSes like Twitter [8, 11, 12, 13, 15], semi-private SNSes like Facebook [9, 12, 13, 14], and limited-membership SNSes such as those in the enterprise [18]. Indeed, questions are among the posts readers consider most interesting in their feeds [1].

Targeting SNS questions to the right set of potential answerers is challenging; indeed, most SNS questions receive no replies [8]. Some technical solutions to optimize the SNS Q&A experience have been explored. Routing-based approaches include expertise-oriented systems like Aardvark [7] or systems that estimate the likelihood of a user's availability for response [11]. Alternatively, socially-embedded search engines have been designed that can respond to users' SNS questions with algorithmically [6] or crowd-generated [8] answers.

However, most social media users don't employ technical solutions to ensure a response; instead, most rely on simpler, broadcast-based strategies, despite the challenges in estimating what subset of one's network is likely to be exposed to a particular post [2]. To optimize for the success of broadcast questions, many users employ linguistic approaches, such as using scoping keywords to signify the subset of one's network ("Do any of my programmer friends know…?") [17], or employing signifiers of social capital ("Thanks in advance…") [9]. Including tags, such as Twitter's hashtags, is another common approach [8].

The use of hashtags in questions can serve dual roles [16, 19], labeling a post's type/topic (i.e., that it is a question in need of reply) as well as flagging it for visibility to communities of users that monitor particular hashtags [3]. This latter role is potentially important for enabling users to truly harness the promise of social media to connect them to "weak ties" (or strangers), who tend to be sources of novel information [5]. However, studies of Facebook suggest that users tend to reinforce bonds with existing connections rather than forging new ones [10], and that strong ties typically form the crux of Facebook Q&A exchanges [14].

In this Note, we investigate how hashtags impact the audience of questions posed on Twitter. This research answers the questions: (1) When users ask questions that employ hashtags, do they use tags to indicate topic, or to reach a tag-following audience?, and (2) What is the actual makeup of the responses for tagged questions, and are users estimating audience accurately? We contribute evidence of a paradox: users believe that their questions will be exposed to not only their followers, but also to a community that monitors Q&A-oriented hashtags; however, nearly all replies actually come from existing connections. This exposes a rich area for future work in exposing valuable answerers to SNS questions from outside users' networks, so as to realize the potential of SNSes to provide better support for Q&A practices, as well as for helping users reach beyond their bubbles to build new social ties.

## UNDERSTANDING Q&A

In order to understand how the dual nature of hashtags as both topical markers and community indicators influences the results of SNS Q&A, we employ a mixed-methods approach, comprising a survey of Twitter users and an analysis of thousands of Q&A exchanges gathered from Twitter's public Firehose stream.

### Surveying Question Askers

We conducted a survey of U.S.-based Twitter users over a one-day period in August 2013 to probe how participants viewed social tagging for SNS questions. We recruited participants using Twitter advertisements, offering a $2 incentive for survey completion. We received 97 responses, of which 46 were discarded as spam for reasons such as providing an invalid Twitter handle. Of the 51 valid respondents, 65% indicated having asked questions on Twitter (Table 1). Of this group, a majority used hashtags in questions. Most anticipated both followers and those who monitor hashtags would see and reply to questions. They cited hashtags as both topic markers and communities. Respondents provided examples of hashtags typically used in their question tweets; these sample tags were about equally split between generic Q&A tags (e.g., *#lazyweb, #help*) and topic-specific tags (e.g., *#songtitle, #cityname*).

### Identifying Q&A Hashtags

To investigate the extent to which hashtags functioned as communities for the purpose of answering questions, we must identify a set of hashtags on which to focus. To build a general set of question-signaling hashtags, we sampled over 60 million public, English-language tweets that ended in question marks from the Twitter Firehose from May 9th to 16th, 2013. We further restricted this dataset by only including tweets with hashtags; from users with a timezone set (a signifier of active accounts); and starting with question words like "who," "where," "when," "why," and "how." This left us with about five million tweets, of which 22.3% received a reply. We identified 2,365 hashtags that were used in at least 50 questions in our dataset.

We then manually inspected this list of hashtags looking for tags that obviously signaled question content. We ultimately produced a list of 40 hashtags. We had 3 different Amazon Mechanical Turk workers rate a random selection of 30 questions from each hashtag, classifying them based on quality and content so that we could understand the sorts of questions that were asked. Using the ratings, manual coding, and Turker qualitative responses, we arrived at a shortlist of 15 hashtags (Table 2) that had reasonable question content and at least 5 tweeted questions a day. In the questions posted to these tags we observed hashtags used as both topical affixes (<question>? #tag1 #tag2) and direct addresses (Hey #tag, <question>?).

| *Who do you expect to see/answer your question?* | |
|---|---|
| My followers: | **85%** |
| People who follow the hashtag I used: | **73%** |
| I don't know: | **12%** |
| ***Use hashtags when posting questions:*** | **66%** |
| ***Why?*** | |
| To reach an audience for the hashtag: | **48%** |
| To indicate a question or the topic: | **58%** |
| Because I saw others do it: | **15%** |
| To indicate what kind of answer I want: | **3%** |
| *How often do you get answers for questions you ask?* | |
| Never: | **6%** |
| Rarely: | **30%** |
| Sometimes: | **39%** |
| Often: | **18%** |
| Always: | **6%** |
| *In general, how satisfying are the answers you get?* | |
| Very dissatisfying: | **3%** |
| Dissatisfying: | **3%** |
| Neutral: | **48%** |
| Satisfying: | **42%** |
| Very satisfying: | **3%** |

**Table 1: Survey responses for the participants who had asked questions on Twitter**

### Observing Dyad Tie Strength

Armed with a set of 15 question-indicating hashtags, we collected a dataset of all questions (subject to our prior filtering criteria) and replies posted over the course of one day (August 2nd, 2013) employing one or more of the 15 hashtags. Out of 5,859,592 filtered questions for the day, 7,253 questions contained our chosen hashtags. Out of all questions, 13.4% received replies, while 34.1% of questions using one or more of our 15 hashtags received replies, significantly more ($c^2(1, N=5,859,592)=2683.0, p<0.001$). This is potentially due to spam content or more rhetorical questions occurring outside of the hashtags we selected.

To evaluate who is actually replying to the questions posted to the different hashtags, we evaluated all of the questions posted to the hashtags that received replies. For each of these questions, we only examined the first reply in order to limit the potential relationship between the replier and questioner. Were we to examine replies beyond the first we would have to consider not only the replier's relationship to the questioner, but also their relationship to other repliers in order to estimate the question's audience. We gathered the followers and friends of each questioner/replier dyad.

| Hashtag | # Q/R Dyads | Percent Connected | Percent Mutual |
|---|---|---|---|
| *General sample* | *400* | *93.9* | *84.0* |
| *Hashtags overall* | *2193* | *94.1* | *85.4* |
| #askingforafriend | 27 | 96.3 | 88.9 |
| #asktwitter | 549 | 94.2 | 89.3 |
| #help | 533 | 90.1 | 80.7 |
| #ineedanswers | 20 | 95.0 | 95.0 |
| #justasking | 42 | 88.1 | 73.4 |
| #lazyweb | 13 | 92.3 | 84.6 |
| #opinions | 11 | 100 | 90.9 |
| #qtna | 133 | 97.0 | 92.5 |
| #question | 31 | 90.3 | 71.0 |
| #questions | 20 | 100 | 80.0 |
| #questionsthatneedanswers | 20 | 90.0 | 80.0 |
| #randomquestion | 18 | 100 | 83.3 |
| #replytweet | 632 | 96.2 | 86.1 |
| #seriousquestion | 100 | 95.0 | 88.0 |
| #twoogle | 44 | 93.2 | 77.3 |

**Table 2: General statistics and tie strengths for a random sample of 400 dyads and the 15 hashtag sample**

While rich models of evaluating tie strength between Twitter users have been proposed [4], a simpler model of connectedness is sufficient to address our research question of whether hashtagged questions are replied to by a user's own network or by a community of strangers comprising a hashtag's audience. Dyads are connected if one of them follows the other. We also computed the proportion of the questioner's follower/friend network that is shared with the replier, giving a gross estimate of tie strength.

Table 2 describes the resulting connectedness for each hashtag. Over all hashtag communities, 93.8% of Q&A dyads were connected to each other. For 85.4% of dyads, the connection was mutual. Questioners shared a mean of 7.3% (SD 10.2%) of their followers with their replier. The distribution of follower sharing was roughly exponential, with a median at 4.1% and quartiles at 1.3% and 9.6%. There were no large differences between hashtag communities in terms of connectedness. We also evaluated tie strength over a random sample of 400 question/reply dyads regardless of hashtag use using the same approach. The connectedness between dyads in this sample and those in the hashtag communities do not differ significantly.

These results suggest that people generally reply to existing connections. While hashtags may function as topical markers, as evidenced by the increased reply rate of questions that used hashtags, it seems that if there are users affiliated with tags, then they are not spending the effort to reply to strangers' questions. It may be that question-indicating hashtags do in fact have an audience of "lurkers" [13], but those users are not willing to take the time to respond, perhaps because of a lack of incentives.

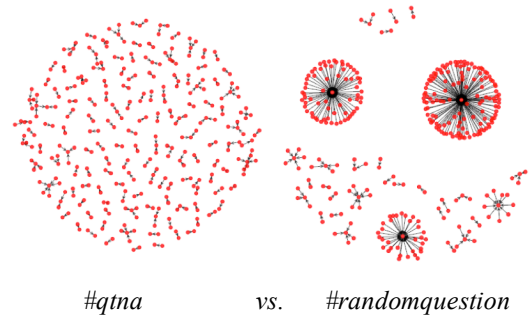

#qtna      vs.      #randomquestion

**Figure 1: Question/reply graph for two hashtag communities. Note the stars in #randomquestion versus the disconnected dyads in #qtna common in the other hashtags.**

**Observing Question/Reply Graph Structure**

To further explore the communities associated with our 15 chosen hashtags, we consider the structural features of the question/reply relationship graph over all replies. Ties in this network indicate user A replied to a question posted by user B. The large majority of the components in this network are isolated dyads; on average across the 15 hashtags, 67.8% of components were of size 2. Only one hashtag, #asktwitter, showed any reciprocal replying. Another interesting finding was the presence of star configurations, indicating individuals who ask lots of questions and received many answers - #randomquestion in particular had a few highly active questioners. This may be indicative of a spammer or well-connected user posting many questions in broadcast. Overall we observe a lack of connection in the network. These structural differences pose rich areas for future investigation.

**DISCUSSION**

For general consumers of content on Twitter, hashtags seem to serve a dual purpose. They indicate the topic of a tweet, and also suggest a group of people who may be interested in that topic [16, 19]. By monitoring the tag, people may discover new friends, sources of information, or interesting anecdotes. However, question-indicating hashtags do not seem to promote such extra-network engagement.

Because answering a question expends effort, the audience of a hashtag may be unwilling to help out if the benefits are not clear. This manifests in the paradox between Twitter users' expectations and reality. In our survey, respondents expected that hashtag followers would see and possibly answer their question. Many employed hashtags when tweeting questions specifically to reach these imagined audiences. However, for general question-signifier tags, these expectations are not realized in practice. We observed that for the 15 most active, non-spam Q&A hashtags, most of the replies a person gets are actually from their existing network connections. The only benefit the hashtag seems to provide is improving the response rate overall, though this may be more indicative of factors such as questioner effort.

Note that while we recruited survey participants from a general U.S. audience on Twitter, our relatively small sample of 51 respondents may not have captured the whole variety of perspectives of Twitter users; practices (and expectations) on non-English-language hashtags may also differ. It is also worth noting that our sample of hashtags was limited to high-activity tags containing English-language questions with interrogative grammar indicators – one might expect different results in more niche or specialized hashtags. For instance, corporately-monitored hashtags may quickly respond to strangers' questions for business reasons, and niche fandoms may be so enthused that they are happy to respond – perhaps it is the very generality of hashtags like #twoogle, which signify a tweet's type rather than a tweet's topic, that precludes the formation of communities. This might explain why our observed overlap is much higher than earlier studies [15]. Our evaluation of tie strength does not capture the full complexity of relationships on Twitter. Foremost, it does not consider anything outside of Twitter or anything beyond one hop on the network, though this would only increase the connectedness. One potential explanation for the high number of connections may be a result of homophily.

The disconnect between potential hashtag audience and who is actually expending effort to respond to posts suggests many avenues for future research. Our findings are largely descriptive, future studies might relate them to existing theories such as reciprocity and social capital. While estimating audience for SNS posts is challenging [2], perhaps users may be able to accurately estimate who in their network will expend effort should they ask them to. Similarly, if followers are more likely to answer, then SNS question-routing systems might be designed so as to emphasize social ties rather than communities of unaffiliated answerers. Such systems might offer (or more clearly emphasize) incentives for answering so as to maximize the benefits of social media for addressing information needs and facilitating tie creation. SNS systems that employ tagging features might display statistics or visualizations that help posters understand the differential characteristics associated with different tags (activity level, spam level, community composition, reply level, etc.).

## CONCLUSION

Users of SNSes like Twitter often use rich social tagging systems not only to mark content topics, but also to access a community of people interested in a given topic. This may be especially useful in the case of social network information-seeking, as those who are interested in a topic or in general Q&A activities may be best able to answer a user's question. Through a survey of Twitter users, we saw that participants indeed employed hashtags when asking questions with an expectation that they were accessing an audience as well as marking topic. However, through studying the relationship between questioner and answerer, we see that almost all replies come from social ties rather than a separate audience. These findings have bearing not only for how we construct and support information-seeking via social networks, but also in how we evaluate audience not only for exposure but also for expertise and will to expend effort on social networking services.

## REFERENCES

1. André, P., Bernstein, M.S., and Luther, K. Who Gives a Tweet? Evaluating Microblog Content Value. *CSCW 2012*.
2. Bernstein, M., Bakshy, E., Burke, M., and Karrer, B. Quantifying the Invisible Audience in Social Networks. *CHI 2013*.
3. Cook, J., Kenthapadi, K., and Mishra, N. Group Chats on Twitter. *WWW 2013*.
4. Gilbert, E. Predicting Tie Strength in a New Medium. *CSCW 2012*.
5. Grannovetter, M. The Strength of Weak Ties. *American Journal of Sociology*, May 1973.
6. Hecht, B., Teevan, J., Morris, M.R., and Liebling, D. SearchBuddies: Bringing Search Engines into the Conversation. *ICWSM 2012*.
7. Horowitz, D. and Kamvar, S.D. The Anatomy of a Large-Scale Social Search Engine. *WWW 2010*.
8. Jeong, J-W., Morris, M.R., Teevan, J., and Liebling, D. A Crowd-Powered Socially Embedded Search Engine. *ICWSM 2013*.
9. Jung, Y., Gray, R., Lampe, C., and Ellison, N. Favors from Facebook friends; Unpacking dimensions of social capital. *CHI 2013*.
10. Lampe, C., Ellison, N., and Steinfield, C. A Face(book) in the crowd: Social searching vs. social browsing. *CSCW 2006*.
11. Mahmud, J., Zhou, M.X., Megiddo, N., Nichols, J., and Drews, C. Recommending targeted strangers from whom to solicit information on social media. *IUI 2013*.
12. Morris, M.R., Teevan, J., and Panovich, K. What Do People Ask Their Social Networks, and Why? A Survey Study of Status Message Q&A Behavior. *CHI 2010*.
13. Morris, M.R. Collaborative Search Revisited. *CSCW 2013*.
14. Panovich, K., Miller, R., and Karger, D. Tie strength in question & answer on social network sites. *CSCW 2012*.
15. Paul, S.A., Hong, L., and Chi, E.H. Is Twitter a good place for asking questions? A characterization study. *ICWSM 2011*.
16. Romero, D.M., Tan, C., and Ugander, J. On the Interplay between Social and Topical Structure. *ICWSM 2013*.
17. Teevan, J., Morris, M.R., and Panovich, K. Factors Affecting Response Quantity, Quality, and Speed for Questions Asked via Social Network Status Messages. *ICWSM 2011*.
18. Thom, J., Helsley, S.Y., Matthews, T.L., Daly, E.M., and Millen, D.R. What Are You Working On? Status Message Q&A within an Enterprise SNS. *ECSCW 2011*.
19. Yang, L., Sun, T., Zhang, M., and Mei, Q. We Know What @You #Tag: Does the Dual Role Affect Hashtag Adoption? *WWW 2012*